\documentclass[twocolumn]{aastex63}

\usepackage[T1]{fontenc}
\usepackage{aas_macros}
\usepackage{natbib}
\usepackage{times}

\newcommand{\package}[1]{\textsl{#1}}

\newcommand{\msun}{\mbox{M$_\odot$}}

\newcommand{\gyr}{\mbox{${\rm Gyr}$}}

\newcommand{\rp}{\mbox{$R_{\rm p}$}}
\newcommand{\re}{\mbox{R$_\oplus$}}

\newcommand{\be}{\begin{equation}}
\newcommand{\ee}{\end{equation}}
\newcommand{\bea}{\begin{eqnarray}}
\newcommand{\eea}{\end{eqnarray}}

\received{2020 October 22}
\revised{2020 November 19}
\accepted{2020 November 22}

\submitjournal{ApJ Letters}
\shorttitle{Stellar Clustering and the Planet Radius Valley}
\shortauthors{Kruijssen, Longmore \& Chevance}

\usepackage{amsmath}

\begin{document}\sloppy\sloppypar\raggedbottom\frenchspacing

\title{\vspace{-9mm}Bridging the Planet Radius Valley: Stellar Clustering as a Key Driver for Turning Sub-Neptunes into Super-Earths}

\correspondingauthor{J.~M.~Diederik~Kruijssen}
\email{kruijssen@uni-heidelberg.de}

\author[0000-0002-8804-0212]{J.~M.~Diederik~Kruijssen}
\affil{Astronomisches Rechen-Institut, Zentrum f\" ur Astronomie der Universit\"at Heidelberg, M\"onchhofstra\ss e 12-14, D-69120 Heidelberg, Germany}

\author[0000-0001-6353-0170]{Steven~N.~Longmore}
\affil{Astrophysics Research Institute, Liverpool John Moores University, IC2, Liverpool Science Park, 146 Brownlow Hill, Liverpool L3 5RF, UK}

\author[0000-0002-5635-5180]{M\'{e}lanie~Chevance}
\affil{Astronomisches Rechen-Institut, Zentrum f\" ur Astronomie der Universit\"at Heidelberg, M\"onchhofstra\ss e 12-14, D-69120 Heidelberg, Germany}


\vspace{-5mm}
\begin{abstract}\noindent
Extrasolar planets with sizes between that of the Earth and Neptune ($\rp=1{-}4~\re$) have a bimodal radius distribution. This `planet radius valley' separates compact, rocky super-Earths ($\rp=1.0{-}1.8~\re$) from larger sub-Neptunes ($\rp=1.8{-}3.5~\re$) hosting a gaseous hydrogen-helium envelope around their rocky core. Various hypotheses for this radius valley have been put forward, which all rely on physics internal to the planetary system: photoevaporation by the host star, long-term mass loss driven by the cooling planetary core, or the transition between two fundamentally different planet formation modes as gas is lost from the protoplanetary disc. Here we report the discovery that the planet radius distribution exhibits a strong dependence on ambient stellar clustering, characterised by measuring the position-velocity phase space density with \textit{Gaia}. When dividing the planet sample into `field' and `overdensity' sub-samples, we find that planetary systems in the field exhibit a statistically significant ($p=5.5\times10^{-3}$) dearth of planets below the radius valley compared to systems in phase space overdensities. This implies that the large-scale stellar environment of a planetary system is a key factor setting the planet radius distribution. We discuss how models for the radius valley might be revised following our findings and conclude that a multi-scale, multi-physics scenario is needed, connecting planet formation and evolution, star and stellar cluster formation, and galaxy evolution.
\end{abstract}

\section{Introduction}
\label{sec:intro}
One of the most intriguing results of recent exoplanetary surveys has been the discovery of a bimodal distribution of planet radii between $R=1{-}4~\re$ \citep{fulton17,fulton18}. By combining the precise planetary radius measurements provided by NASA's \textit{Kepler} mission \citep{borucki10,borucki11}, the exquisite parallax measurements of ESA's \textit{Gaia} satellite \citep{gaia16,gaia18}, and astroseismic measurements, recent work has been able to unambiguously establish the existence of this planet `radius valley' \citep{vaneylen18,berger18,berger20,hardegree20}. A population of compact, rocky super-Earths ($\rp=1.0{-}1.8~\re$) is separated from a population of larger sub-Neptunes ($\rp=1.8{-}3.5~\re$) hosting a gaseous hydrogen-helium envelope around their rocky core. In between both populations ($\rp=1.5{-}2.0~\re$), there is a factor-of-two deficit of planets. The precise location of the radius valley decreases weakly with the orbital period and increases weakly with the host stellar mass \citep{fulton18}. The incidence ratio of super-Earths to sub-Neptunes also carries a weak dependence on the stellar system age, increasing from $0.61\pm0.09$ for ages $<1~\gyr$ to $1.00\pm0.10$ for ages $>1~\gyr$ \citep{berger20}.

Several physical mechanisms have been put forward to explain the existence of the radius valley and its dependence on the planet-host system properties. In many recent works, the radius valley is interpreted as evidence that super-Earths form by shedding sub-Neptunes of their gaseous envelopes. However, the physical mechanism responsible is debated.

Originally, the radius valley had been predicted by models describing the photoevaporation of planetary envelopes by the X-ray and extreme ultraviolet (EUV) irradiation from the host star \citep{owen13,owen17,lopez13,jin14}. Motivated by its observational discovery, a variety of subsequent, more detailed models have quantified this scenario further \citep[e.g.][]{mordasini20,rogers20}. Key features of these models are that most of the evolution takes place in the first 100~Myr and that the radius valley shifts to larger radii for stronger irradiation (shorter orbital periods), where a more massive rocky core is required to retain a gaseous envelope.

As an alternative explanation, a series of recent papers has proposed that sub-Neptune planets can lose their gaseous envelopes by core-powered envelope mass loss through a Parker wind \citep{ginzburg18,gupta19,gupta20}. In this scenario, the cooling luminosity of a rocky planetary core can erode a low-mass gaseous envelope, but would leave a massive one unaffected. Key features of this model are that the envelope loss takes place on Gyr timescales and that the radius valley shifts to larger radii for higher envelope equilibrium temperatures (shorter orbital periods), because the envelope density and sound speed at the Bondi radius increase with temperature, such that the mass loss rate increases too \citep{ginzburg16}.

Finally, it has been proposed that the small-radius, high-density population of super-Earths may have at least partially formed without a gaseous envelope, i.e.\ they never were sub-Neptunes \citep{lopez18}. In this scenario, the protoplanetary disc loses its gas through evaporation, such that at late times ($>10$~Myr) only rocky planets can form \citep[also see][]{lee14,lee16}. This mechanism could plausibly act in conjunction with photoevaporation, core-powered mass loss, or both \citep{lee20}.

In the absence of any further empirical input, future observations are expected to help distinguish between these different scenarios. For instance, the photoevaporation and core-powered mass loss mechanisms predict a different dependence of the radius valley on the system age \citep{berger20,gupta20,rogers20}, and the late rocky planet formation scenario predicts a dependence on the orbital period that differs from both other scenarios \citep{lopez18}. All of the discussed mechanisms may be expected to play a role to some degree \citep{lee20}. The key question going forward is what their relative contributions are to the observed radius valley.

In this \textit{Letter}, we add a new and important empirical dependence to the discussion. We take the sample of known exoplanets with radii $\rp=1{-}4~\re$ and orbital periods $P=1{-}100$~days and use the ambient stellar phase space density obtained with \textit{Gaia} to divide the sample into low and high ambient stellar phase space densities \citep{wklc20}, which we refer to as planets residing in the `field' and in `overdensities', respectively. This allows us to assess the potential impact of stellar clustering, e.g.\ through external photoevaporation or dynamical perturbations. We find that planetary systems in the field exhibit a statistically significant ($p=5.5\times10^{-3}$) dearth of planets below the radius valley compared to systems in phase space overdensities. This result mirrors our findings in a set of companion papers, where we investigate the impact of stellar clustering on the orbital period distribution of planets and the incidence of hot Jupiters \citep{wklc20}, as well as orbital resonances and planetary multiplicity \citep[i.e.\ the `Kepler dichotomy', \citealt{lissauer11}]{longmore20}, and the correlation between the properties of adjacent planets \citep[i.e.\ `peas in a pod', \citealt{weiss18}]{chevance20e}. 

Our findings imply that the distribution of planets around the radius valley depends on the degree of stellar clustering in the large-scale environment of a planetary system. At a minimum, this discovery requires modifications to each of the scenarios proposed so far that aim to explain the radius valley. At its extreme, it could require a new scenario that naturally incorporates an environmental dependence of the radius valley. After presenting the results of our analysis, we conclude this \textit{Letter} with a discussion of how models for the radius valley might be revised following our findings.

\section{Observational data}
\label{sec:data}
\subsection{Quantification of stellar clustering}
To assess how the radius valley might depend on ambient stellar clustering, we adopt the classification of observed planetary systems into field and overdensity systems by \citet{wklc20}. The procedure for constructing the classification is as follows. The relative position-velocity phase space densities are calculated for all known exoplanet host stars in the \citet{exoarchive} that have radial velocities from \text{Gaia}'s second data release \citep{gaia18}. At the time of sample construction (May 2020), this restricts the sample to 1525 out of 4141 confirmed exoplanets. The same relative phase space density is calculated for all neighbouring stars within a three-dimensional distance of 40~pc from the exoplanet host star. We then define the probability $P_{\rm null}$ that the resulting distribution of phase space densities of the stars within 40~pc of the planetary system is described by a unimodal Gaussian distribution. For the phase space density distributions that are not well described by a single log-normal, we carry out a double-lognormal decomposition and subsequently assign each host star a probability of belonging to the low-density component ($P_{\rm low}$) or the high-density component ($P_{\rm high}\equiv1-P_{\rm low}$). In the following, we define the \textit{field} sample as exoplanet host stars with $P_{\rm low}>0.84$ and the \textit{overdensity} sample as stars with $P_{\rm high}>0.84$. Exoplanet host stars with less than 400 neighbours within 40~pc, $P_{\rm null}\geq0.05$, or $0.16\leq P_{\rm low/high}\leq0.84$ are considered to have an ambiguous phase space density classification and are removed from the sample, leaving a total of 1033 planets.

\begin{figure*}
\includegraphics[width=\hsize]{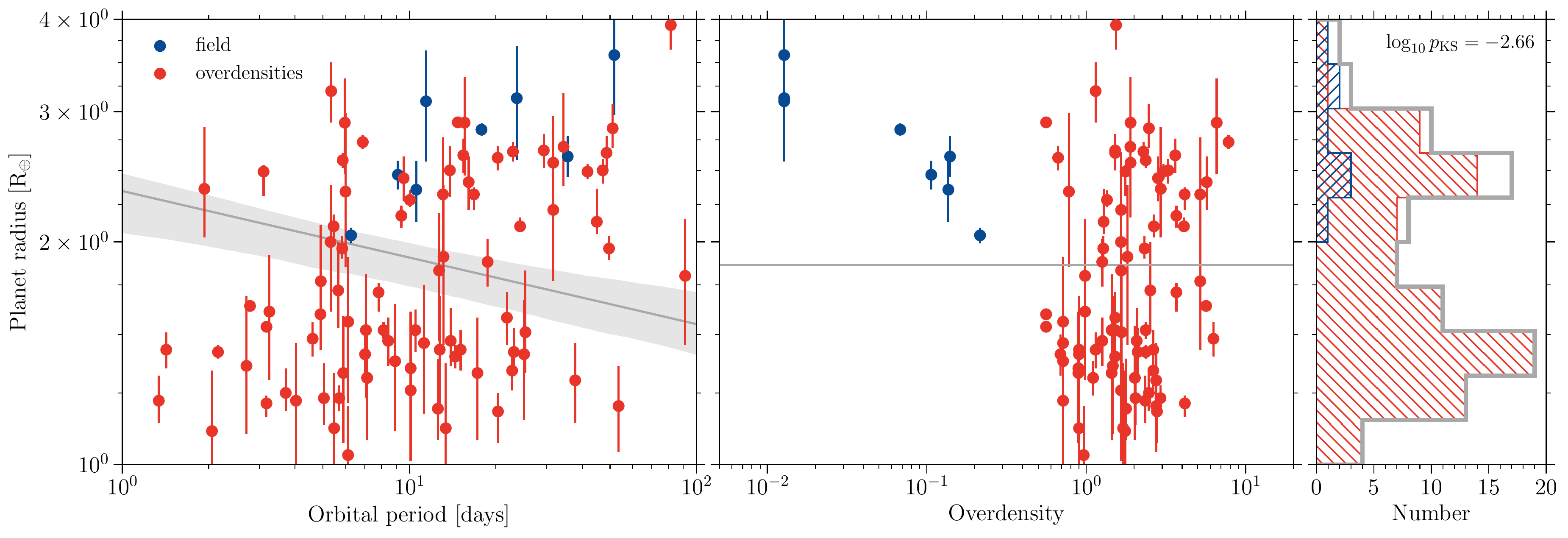}%
\caption{
\label{fig:radvalley}
Distribution of planet radii as a function of their orbital period (left) and of the six-dimensional phase space overdensity of the host star relative to its neighbouring stars within a 40~pc radius (middle). Planets orbiting field stars are marked in blue, whereas planets orbiting stars in overdensities are marked in red. In the left panel, the grey line indicates the observed location of the radius valley (\autoref{eq:radvalley}; taken from \citealt{vaneylen18}), with the shaded area indicating the 16th-84th percentile range given the uncertainties on the location of the valley. In the middle panel, the grey line indicates a constant radius, even if the valley may depend on the phase space overdensity. The histograms in the right panel illustrate how the one-dimensional planet radius distribution differs between both samples, with a KS test $p$-value of $2.2\times10^{-3}$. This figure shows that all planets orbiting field stars reside above the radius valley, i.e.\ all of them are sub-Neptunes and none are super-Earths.
}
\end{figure*}

\subsection{Planet sample}
The sample of planetary systems with a phase space density classification is restricted further using the following sample cuts. Systems with ages younger than 1~Gyr are omitted to exclude planetary systems that might not yet have stabilised \citep[e.g.][]{kennedy13}. Systems older than 4.5~Gyr are excluded, because the occurrence rate of overdensities drops precipitously at older ages, which is plausibly caused by their dynamical dispersal \citep{wklc20}. As quantified in \S\ref{sec:results}, the results do not change when varying the maximum age in the range 4--5~Gyr. We additionally restrict the host stellar masses to a narrow interval of $M=0.7{-}2.0~\msun$ to avoid indirectly probing any dependence on the host mass. Finally, we select the planet sample to be constituted by super-Earths and sub-Neptunes with radii $\rp=1{-}4~\re$ and orbital periods $P=1{-}100$~days. In doing so, we restrict ourselves to planets for which the radii have been measured directly rather than being derived from masses through a planet mass-radius relation. This leaves a final sample of 94 planets, with 8 planets residing in the field and 86 residing in overdensities.

\section{Environmental dependence of\break the planet radius valley}
\label{sec:results}
In \autoref{fig:radvalley}, we show the distribution of planet radii as a function of the orbital period, dividing the planet sample into field and overdensity systems. The well-known dearth of planets around $\rp=1.5{-}2~\re$ is clearly visible, with a substantial population of planets above and below this radius valley. Strikingly, no \textit{field} planets are found below the radius valley, i.e.\ none of them have radii $\rp<2~\re$. By contrast, more than half of the planets in phase space overdensities fall below the radius valley. Given that our sample only includes 8 field planets, the different radius distributions could plausibly result from small-number statistics. Nonetheless, the contrast is intriguing, because it suggests that the large-scale stellar environment might play an important role in driving planets across the radius valley.

We verify whether the difference in radius distribution between our sub-samples might result from covariance with another, possibly more fundamental observable. To do so, we divide the sample into planets in the field, planets in overdensities above the radius valley, and planets in overdensities below the radius valley. The 16th, 50th, and 84th percentiles of various other observables are shown for each of these sub-samples in Table~\ref{tab:hosts}. No significant differences are found in host stellar mass, metallicity, and age, implying that any variation in these quantities cannot explain the difference in planet radius distribution between overdensities and the field. The distance from the Solar System might exhibit a systematic offset, but the standard deviations are considerable. The distance offset simply arises from the specific form of the phase space density structure in the solar neighbourhood and mirrors the difference in distance distributions found in \citet{wklc20}. However, the distances differ in the opposite sense of what could potentially have explained the radius-period distribution shown in \autoref{fig:radvalley}. Field systems reside at closer distances from the Solar System on average, which would enable the detection of smaller planets, but their planet radii all reside above the radius valley. In summary, we find that the difference between overdensities and the field is not caused by covariance with other quantities.

\begin{deluxetable*}{l @{\hspace{0.5cm}} l l l l}
\tablehead{
Planet sub-sample & Stellar mass & Stellar metallicity & System age & Distance \\
& [$\msun$] & [dex] & [Gyr] & [pc]
}
\decimals
\setlength{\tabcolsep}{3pt}
\startdata
Field (all above valley) & $1.01_{-0.20}^{+0.07}$ & $-0.08_{-0.12}^{+0.15}$ & $3.7_{-0.9}^{+0.1}$ & $182_{-159}^{+223}$ \\
Overdensities above valley & $1.10_{-0.24}^{+0.17}$ & $0.05_{-0.11}^{+0.16}$ & $3.1_{-1.1}^{+1.2}$ & $306_{-81}^{+224}$ \\
Overdensities below valley & $1.14_{-0.20}^{+0.11}$ & $0.03_{-0.06}^{+0.13}$ & $3.0_{-0.7}^{+1.0}$ & $401_{-151}^{+179}$ \\
\enddata
\caption{Median host stellar properties for each of the three planet sub-samples listed in the first column. Uncertainties indicate the 16th and 84th percentiles of the distributions.\vspace{-5mm}
}
\label{tab:hosts}
\end{deluxetable*}

To quantify the statistical significance of the observed difference between the radius-period distributions of planets in the field and those in overdensities, we carry out a simple statistical experiment. The goal of this experiment is to determine the probability of randomly drawing $n$ field planets from the entire sample shown in \autoref{fig:radvalley}, of which $k$ fall above the radius valley, i.e.\ are sub-Neptunes. If the measurements do not have any uncertainties, this is simply given by the binomial distribution:
\be
\label{eq:binom}
P(X=k)=\binom{n}{k}p^k(1-p)^{n-k} ,
\ee
where $p\equiv N_{\rm above}/N_{\rm tot}$ is the probability of drawing a single planet above the radius valley (i.e.\ sub-Neptune), defined as the ratio between the number of planets above the radius valley and the total number of planets. Here, we have defined the radius valley using the expression obtained by \citet{vaneylen18} for a sample of 117 planets orbiting stars characterised to high precision using asteroseismology:
\be
\label{eq:radvalley}
\log{(\rp/\re)}=m\log_{10}{(P/{\rm days})}+a ,
\ee
where $P$ indicates the orbital period, and the best-fitting coefficients are $m=-0.09^{+0.02}_{-0.04}$ and $a=0.37^{+0.04}_{-0.02}$. Substituting the appropriate values $\{k; n, p\}=\{8; 8, 0.457\}$, we obtain $P(X=8)=1.9\times10^{-3}$. This is similar to the result of a two-tailed KS test performed on both radius distributions ($p_{\rm KS}=2.2\times10^{-3}$) and indicates that the observed difference is approximately a $3\sigma$ result.

However, both the observed radii and the definition of the radius valley carry uncertainties, which increase the probability that they might scatter across the radius valley. To incorporate these uncertainties into our statistical assessment, we carry out a Monte-Carlo experiment and generate 100,000 random realisations of the planet sample by drawing each data point from a normal distribution, of which the median and standard deviation are set by the measurement and its uncertainty. For each realisation, we draw $n=8$ random planets and count how many of these reside above the radius valley. The realisations also account for the uncertainty on the location of the radius valley itself by each using coefficients in \autoref{eq:radvalley} that are randomly drawn from a normal distribution set by their best-fitting values and uncertainties. The probability of drawing $k$ planets above the radius valley then follows as the fraction of Monte-Carlo realisations in which $k$ such sub-Neptunes are found.

\begin{figure}
\includegraphics[width=\hsize]{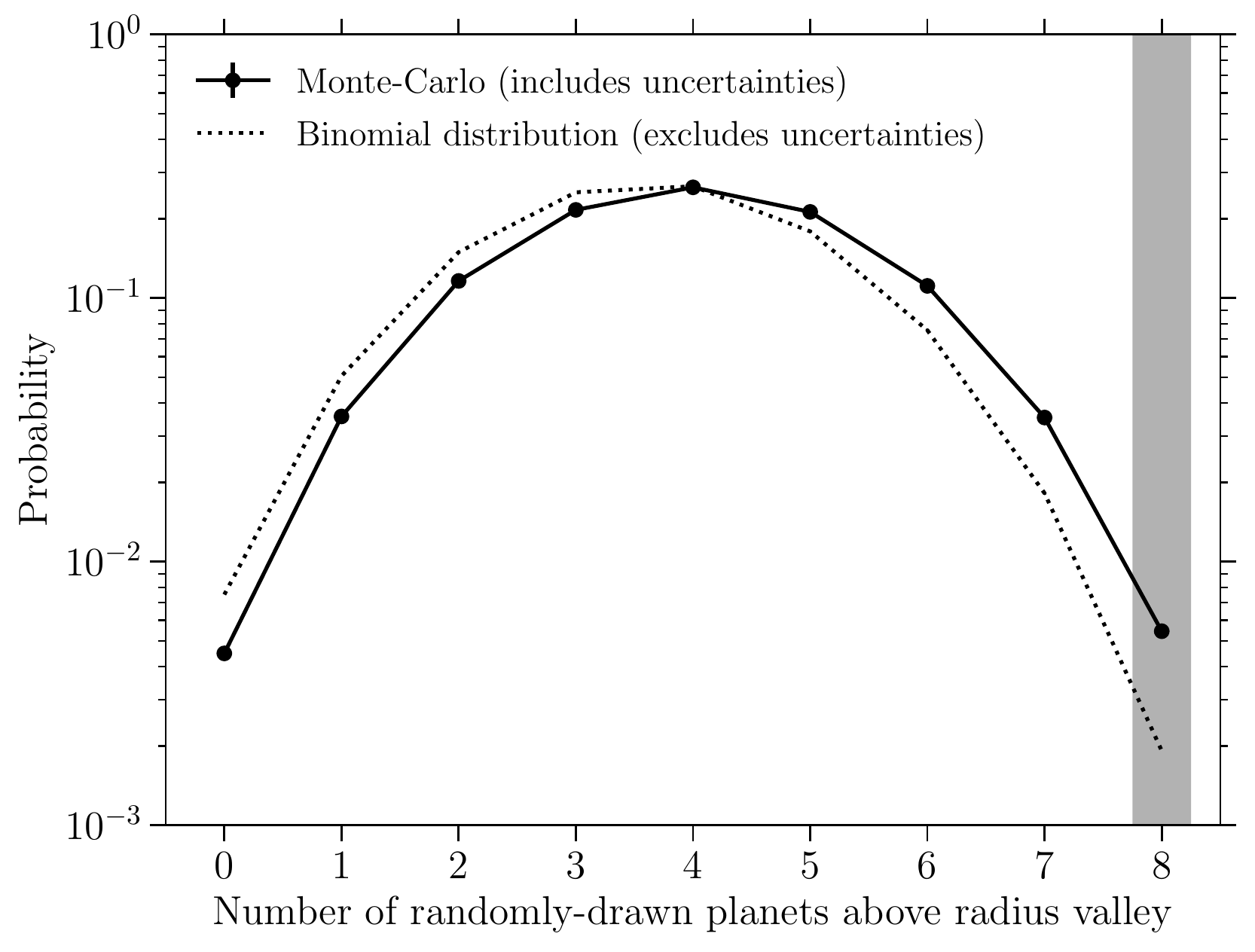}%
\caption{
\label{fig:prob}
Probability distribution of the number of planets residing above the radius valley when randomly drawing 8 `field' planets from the full sample shown in \autoref{fig:radvalley}. The dotted line shows the simple binomial result from \autoref{eq:binom}, whereas the solid line shows the result of the Monte-Carlo experiment described in the text, which accounts for the uncertainties on the planet radii and the location of the radius valley. The observation that 8 field planets reside above the radius valley is marked as a vertical grey band. This figure shows that our observation is unlikely ($P=5.5\times10^{-3}$) to result from random chance.
}
\end{figure}
\autoref{fig:prob} shows the probability distribution of the number of planets above the radius valley ($k$) when drawing $n=8$ planets, both for the binomial expression from \autoref{eq:binom} and the Monte-Carlo experiment described above. As expected, the probability that all 8 drawn planets are sub-Neptunes and reside above the radius valley is higher for the Monte-Carlo experiment than for the simple binomial estimate. Most of this shift is driven by the uncertainties on the coefficients in \autoref{eq:radvalley}. Despite this increase, the probability that all 8 planets orbiting field stars would reside above the radius valley remains small, with $P(X=8)=5.5\times10^{-3}$.\footnote{When changing the maximum system age to $\{4,5\}$~Gyr instead of the adopted 4.5~Gyr, the number of field planets becomes $n=\{7,12\}$, of which $k=\{7,11\}$ reside above the radius valley, with associated probabilities of $\{8.8,5.1\}\times10^{-3}$. Likewise, when omitting the adopted minimum age cut of 1~Gyr, we retain $\{k; n\}=\{8,8\}$ and find a probability of $6.4\times10^{-3}$. This means that the specific choice of the system age range does not affect the results. Going beyond 5~Gyr would be incorrect, because at these old ages the number of systems in overdensities approaches zero due to dynamical dispersal \citep{wklc20}. Therefore, the field sample older than 5~Gyr is contaminated by former overdensity systems.} In other words, the observation that field planets preferentially reside above the radius valley is unlikely to result from random chance. Instead, the preferred interpretation is that stellar clustering might play an important role in driving sub-Neptunes below the radius valley.

\section{Discussion}
\label{sec:disc}
We have divided the observed planet population with radii $\rp=1{-}4~\re$ and orbital periods $P=1{-}100$~days into `field' and `overdensity' samples, based on the six-dimensional stellar phase space density distribution in their vicinity, as observed with \textit{Gaia} \citep{wklc20}. After making this division and applying the necessary sample cuts, we are left with 8 planets orbiting field stars and 86 planets orbiting stars in phase space overdensities. Even though the sample remains small at present, we obtain the intriguing result that none of the field planets reside below the planet `radius valley' at $\rp=1.5{-}2.0~\re$ \citep{fulton17}, which separates super-Earths ($\rp=1.0{-}1.8~\re$) and sub-Neptunes ($\rp=1.8{-}3.5~\re$). We demonstrate that this result is unlikely to result from random chance.\footnote{Of course, these statistics will change with future planet discoveries or phase space density measurements. To decrease the significance of our finding from $P=5.5\times10^{-3}$ (or $3\sigma$) to $P\approx5\times10^{-2}$ (or $2\sigma$), the next two newly-discovered field planets that satisfy our sample cuts would need to be super-Earths. However, if these turn out to be sub-Neptunes, they would strengthen our result further to $P\approx1\times10^{-3}$.} Instead, the preferred interpretation is that the difference in planet radii between overdensities and the field is physical in nature. This means that the large-scale stellar environment might play an important role in turning sub-Neptunes into super-Earths. We now place this result in the context of the observed trends and of the physical mechanisms put forward to date, with the goal of assessing which physical ingredients are needed to explain the existence of the radius valley.

\subsection{Individual scenarios for explaining the radius valley}
\begin{figure}
\includegraphics[width=\hsize]{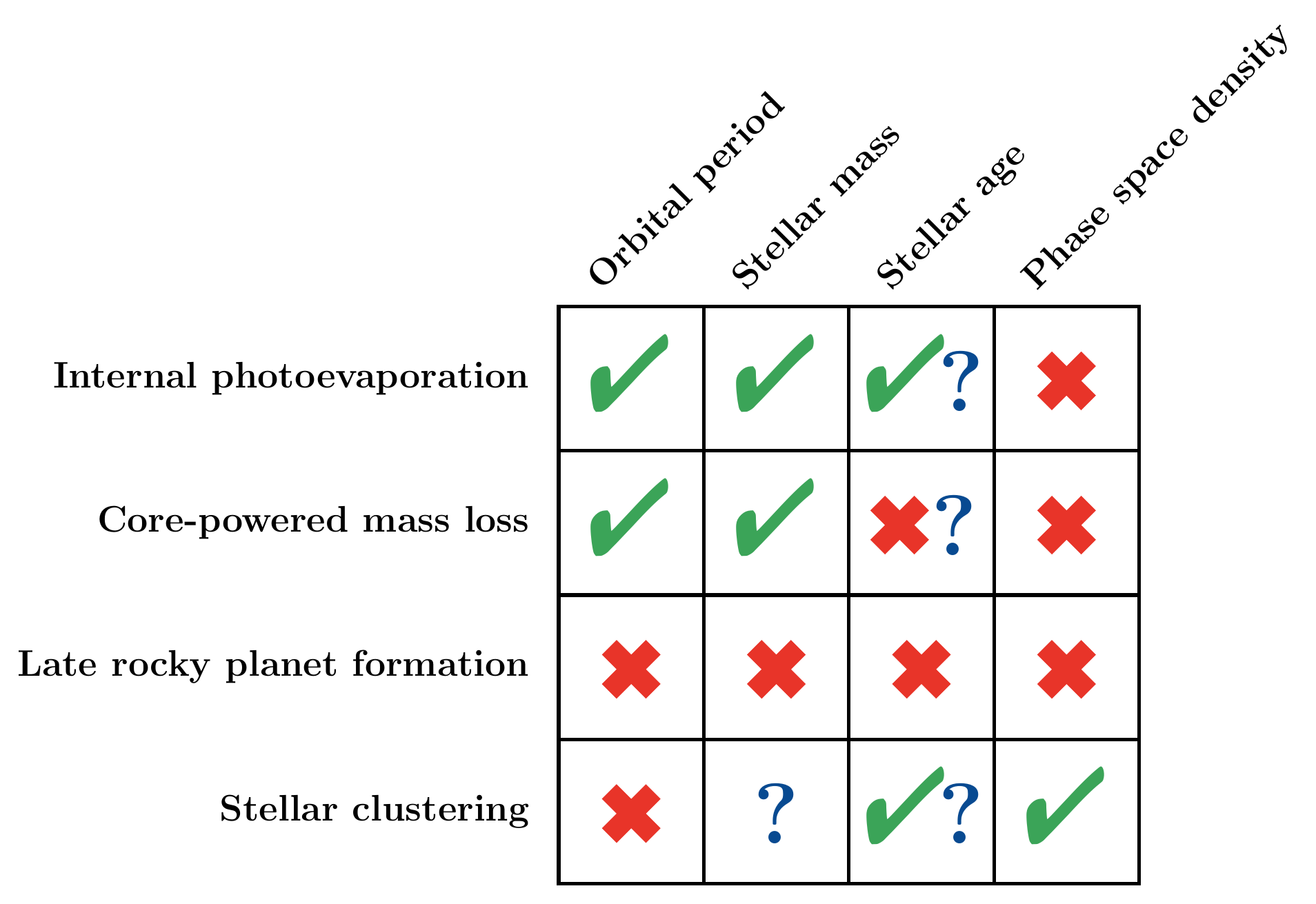}%
\caption{
\label{fig:table}
Summary of the physical mechanisms proposed to explain the radius valley (rows) compared to the observed dependences of the radius valley (columns). A red cross indicates an inconsistency between an observed trend and a single mechanism, a green check mark indicates that the observed trend is qualitatively consistent with the prediction of a single mechanism, and a blue question mark indicates that the mechanism needs to be explored further to unambiguously predict a trend. A combination of a cross or check mark with a question mark indicates a tentative inconsistency or agreement between a mechanism and an observation. Based on this summary, a combination of mechanisms is needed, which at least includes internal photoevaporation and stellar clustering.
}
\end{figure}

We consider the three scenarios outlined in \S\ref{sec:intro} (\textit{internal} photoevaporation by the host star, core-powered mass loss, and late rocky planet formation in a gas-poor disc), and also include `stellar clustering', by which we refer to any influence of the large-scale stellar environment, be it \textit{external} photoevaporation by other stars or dynamical perturbations of the system. In principle, each of these scenarios or mechanisms could lead to the existence of a radius valley, either by atmospheric mass loss (internal photoevaporation, core-powered mass loss, and external photoevaporation due to stellar clustering) or through the coexistence of two separate formation channels (late rocky planet formation). To distinguish between these scenarios, we should therefore focus on how the properties of the radius valley are expected to vary in each of them. For each of these scenarios, we evaluate their (potential) ability to reproduce the observed dependence of the existence or location of the radius valley on the orbital period, host stellar mass, host stellar age, or ambient stellar phase space density. A summary of this assessment is shown in \autoref{fig:table}, and we discuss each element below.

Observationally, the location of the radius valley decreases with orbital period \citep{vaneylen18}. This feature and its slope had been predicted by internal photoevaporation models before the discovery of the radius valley \citep{owen13,lopez13,jin14}, simply because planets orbiting at larger semi-major axes are less irradiated and undergo less atmospheric mass loss. A similar dependence is found in core-powered mass loss models \citep{ginzburg18,gupta19}, because the atmosphere equilibrium temperature and the corresponding mass loss rate both decrease towards larger semi-major axes. In late rocky planet formation models, the orbital period dependence has the opposite sign \citep{lopez18}, because the maximum rocky planet mass (set by the mass within the Hill radius) increases outwards. Finally, stellar clustering is expected to have a weak effect on the orbital period dependence. External photoevaporation should have a similar effect on all planets, independently of their periods, under the reasonable assumption that the distance to the photoevaporating source is much larger than the semi-major axis. Encounters with other stars could rearrange the planetary system, either causing a random redistribution of orbits (which would erase the period dependence) or driving a homologous evolution of the planetary system (which could change the slope of the period dependence). While it is clear that stellar clustering on its own is not responsible for the period dependence, its potential influence on the other mechanisms is an important area for future investigations.

In addition, the location of the radius valley increases with host stellar mass \citep{fulton18}. In the internal photoevaporation model, this is the direct result of stronger irradiation from more massive stars being able to evaporate the envelopes of more massive planetary cores \citep{owen17}. The core-powered mass loss scenario predicts a similar mass dependence \citep{gupta20}, which in that model arises due to a higher equilibrium temperature for planets orbiting more massive stars, leading to correspondingly more atmospheric mass loss. As for the orbital period dependence, the prediction for late rocky planet formation is opposite to the observed trend, because more massive stars imply smaller Hill radii and correspondingly smaller rocky planet masses. For stellar clustering, this time there is a more concrete prediction to be made -- the impact of stellar encounters is predicted to be stronger towards higher host masses, for two reasons \citep{winter20}. Firstly, gravitational focusing increases the encounter rate of more massive stars. Secondly, massive stars with a protoplanetary disc have a larger gravitational radius \citep{hollenbach94} at which the thermal disc wind is launched by internal photoevaporation, making the remaining disc less gravitationally bound to the host star and more susceptible to dynamical, outside-in truncation. This may lead to accelerated disc dispersal, in which case planet formation would take place in a more gas-poor environment. Taken together, these trends could potentially lead to a positive correlation between the radius valley and stellar mass.

A recent study by \citet{berger20} demonstrates that the distribution of planets around the radius valley depends on the system age. The number ratio of super-Earths and sub-Neptunes, increases from $0.61\pm0.09$ for ages $<1~\gyr$ to $1.00\pm0.10$ for ages $>1~\gyr$. This is interesting, because the various scenarios aimed at explaining the radius valley make quite different predictions. Internal photoevaporation operates mostly within the first $100$~Myr, but continues to act afterwards, such that it predicts a ratio $0.77\pm0.09$ for ages $<1~\gyr$ and $0.95\pm0.08$ for ages $>1~\gyr$ \citep[sect.~4.6 and fig.~15]{rogers20}. This trend agrees qualitatively with the observed ratios, but quantitatively shows too little evolution. By contrast, core-powered mass loss acts mostly on $0.5{-}2$~Gyr timescales, such that qualitatively a time evolution is predicted. However, contrary to internal photoevaporation, this time evolution may come too late (or be insufficient) due to the long associated timescale. The number ratio of super-Earths and sub-Neptunes predicted for core-powered mass loss is $\{0.06, 0.52, 0.59\}$ at ages of $\{0.5,3,10\}~\gyr$ \citep[fig.~10]{gupta20}. In contrast to the other mechanisms, late rocky planet formation takes place entirely during the protoplanetary disc lifetime ($\lesssim10$~Myr) and cannot reproduce any age trend on Gyr timescales \citep[e.g.][]{lee14}. Therefore, it cannot be solely responsible for the properties of the radius valley. Finally, phenomenologically speaking, stellar clustering is consistent with an age dependence. Irrespectively of the exact environmental mechanism (stellar dynamical encounters or the external photoevaporation of sub-Neptune gaseous envelopes), stellar clustering most likely manifests itself through a Poisson process, of which the impact naturally increases with age.

\subsection{The need for a multi-scale, multi-physics scenario}
None of the three scenarios that are internal to a planetary system can by themselves explain the observed dependence of the radius valley on the ambient stellar phase space density. Given that a correlation between phase space density and host star properties is ruled out (see Table~\ref{tab:hosts} and \citealt{wklc20}), it is clear that this dependence must be caused by mechanisms related to stellar clustering, such as external photoevaporation or stellar encounters. However, stellar clustering on its own is unlikely to explain some of the other trends in \autoref{fig:table}, such that a combination of mechanisms is necessary to explain the observed properties of the radius valley. In this case, stellar clustering might affect the effectiveness or way in which the internal mechanisms act.

The impact of stellar clustering on each of the internal scenarios differs, such that a preferred combination of mechanisms might be obtained. Core-powered mass loss is a strictly internal process (governed by the planet's internal heat source and the equilibrium temperature set by the host star), making it unlikely that it could be affected by stellar clustering. As a result, it is extremely challenging to reconcile core-powered mass loss with the phase space density dependence of the radius valley identified here. By contrast, it might be possible that stellar clustering accelerates the dispersal of the gas in the protoplanetary disc through external photoevaporation or stellar encounters \citep{winter20}. This would increase the dust-to-gas ratio, promote late rocky planet formation, and increase the impact of internal photoevaporation on sub-Neptune atmospheres by helping clear the disc. Alternatively, stellar clustering might increase the stellar encounter rate long after disc dispersal, leading to regular perturbations of the planetary system, possibly changing the orbital configuration, and causing transient peaks of the internal photoevaporation rate. In either of these two scenarios, the radius valley results from a combination of internal photoevaporation and stellar clustering, with a possible role for late rocky planet formation in the first scenario.

A comprehensive quantitative model for the role of stellar clustering in shaping the planet radius valley would require determining the physical origin of the identified stellar phase space overdensities. Are the overdensities a relic of the clustered formation environment in which most stars form \citep{kruijssen12d,hopkins13d}, or can they be generated at a later time? Both options are not without difficulties. The overdensities arising from phase space clustering at birth may disperse on sub-Gyr timescales \citep{krumholz19,webb20}, much shorter than the ages considered here. This problem could be alleviated by the prediction that the destabilising effects of encounters persist well after a planetary system has escaped the birth environment \citep[e.g.][]{cai18}. Alternatively, the overdensities identified by \citet{wklc20} might have formed later, which is supported by their similarity to the \textit{Gaia} phase space overdensities related to galactic-dynamical processes \citep{quillen18,fragkoudi19}. This would rule out the first of our two suggested combined scenarios, because the present phase space overdensities would be unrelated to processes taking place at the time of planet formation. However, the problem would then be that the encounter rate with field stars is likely too low to explain the correlation between the radius valley and stellar clustering. Even though the physical origin of the phase space overdensities is currently still unknown, it is clear that their characterisation represents a key step towards understanding the origin of the planet radius valley.

Finally, the difference in radius distribution between overdensities and the field has implications for the incidence of Earth-like planets in the habitable zone \citep[$\eta_\oplus$, e.g.][]{kopparapu13}. The individual scenarios for the radius valley that have been put forward previously already predict that $\eta_\oplus$ may not be set at birth, but evolves with age. Our findings now imply that this evolution very likely depends on the clustering of stars in the large-scale stellar environment, which itself is set by galaxy formation and evolution \citep[e.g.][]{pfeffer18}. This means that $\eta_\oplus$ (and cosmic habitability in general) is governed by physics ranging from AU to Mpc scales (\citealt{kruijssen20}; in prep.).

In summary, we find that a combination of physical mechanisms is needed to explain the planet radius valley. This likely requires internal evaporation by the host star and some form of external process driven by stellar clustering (such as external photoevaporation or stellar dynamical encounters), with possibly a role for late rocky planet formation. Core-powered mass loss models face a serious challenge in reproducing the observed dependence of the radius valley on the ambient stellar phase space density. While core-powered mass loss may contribute to the conversion of sub-Neptunes into super-Earths, it cannot dominate this process and explain the radius valley on its own. Irrespectively of the specific combination of processes leading to the radius valley, its dependence on the degree of stellar clustering in the large-scale environment of the planetary system means that it can only be fully understood through a multi-scale and multi-physics approach, connecting planet formation and evolution, star and stellar cluster formation, and galaxy evolution.

\acknowledgments
J.M.D.K.\ and M.C.\ gratefully acknowledge funding from the Deutsche Forschungsgemeinschaft (DFG, German Research Foundation) through an Emmy Noether Research Group (grant number KR4801/1-1) and the DFG Sachbeihilfe (grant number KR4801/2-1). J.M.D.K.\ gratefully acknowledges funding from the European Research Council (ERC) under the European Union's Horizon 2020 research and innovation programme via the ERC Starting Grant MUSTANG (grant agreement number 714907). This research made use of data from the European Space Agency mission \textit{Gaia} (\href{http://www.cosmos.esa.int/gaia}{http://www.cosmos.esa.int/gaia}), processed by the \textit{Gaia} Data Processing and Analysis Consortium (DPAC, \href{http://www.cosmos.esa.int/web/gaia/dpac/consortium}{http://www.cosmos.esa.int/web/gaia/dpac/consortium}). Funding for the DPAC has been provided by national institutions, in particular the institutions participating in the \textit{Gaia} Multilateral Agreement. This research has made use of the NASA Exoplanet Archive, which is operated by the California Institute of Technology, under contract with the National Aeronautics and Space Administration under the Exoplanet Exploration Program.

\software{
\package{matplotlib} \citep{hunter07},
\package{numpy} \citep{vanderwalt11},
\package{pandas} \citep{reback20},
\package{scipy} \citep{jones01},
\package{seaborn} \citep{waskom20}
}

\bibliographystyle{aasjournal}
\bibliography{mybib}

\end{document}